\documentclass[
a4paper
]{ceurart}

\usepackage{listings}
\begin{document}

\copyrightyear{2025}
\copyrightclause{Copyright for this paper by its authors.
  Use permitted under Creative Commons License Attribution 4.0
  International (CC BY 4.0).}

\conference{LAK'25 Workshop: LLMs for Qualitative Analysis in Education
  March 04, 2025, Dublin, Ireland}

\title{Toward Automated Qualitative Analysis: Leveraging Large Language Models for Tutoring Dialogue Evaluation}

% \tnotetext[1]{You can use this document as the template for preparing your publication. We recommend using the latest version of the ceurart style.}


% Megan Gu, Chloe Qianhui Zhao, Claire Liu, Nikhil Patel, Jahnvi Shah, Jionghao Lin, Kenneth R. Koedinger
% Carnegie Mellon University
% {megangu,qianhuiz,claireli,nikhilp,jahnvis,jionghal,kk1u}@andrew.cmu.edu

\author[1]{Megan Gu}[
email=megangu@andrew.cmu.edu
]

\author[1]{Chloe Qianhui Zhao}[
email=cqzhao@cmu.edu
]

\author[1]{Claire Liu}[
email=claireli@andrew.cmu.edu
]

\author[1]{Nikhil Patel}[
email=nikhil@andrew.cmu.edu
]

\author[1]{Jahnvi Shah}[
email=jahnvis@andrew.cmu.edu
]

\author[2, 1, 3]{Jionghao Lin}[
email=jionghao@hku.hk
]
\cormark[1]

\author[1]{Kenneth R. Koedinger}[
email=kk1u@andrew.cmu.edu
]

% \author[1,2]{Dmitry S. Kulyabov}[%
% orcid=0000-0002-0877-7063,
% email=kulyabov-ds@rudn.ru,
% url=https://yamadharma.github.io/,
% ]

\address[1]{Carnegie Mellon University,
  5000 Forbes Ave, Pittsburgh, PA, 15213, United States}

\address[2]{The University of Hong Kong, 
Pokfulam Rd, Hong Kong, China}

\address[3]{Monash University,
Wellington Rd, Clayton VIC 3800, Australia}

% \author[3]{Ilaria Tiddi}[%
% orcid=0000-0001-7116-9338,
% email=i.tiddi@vu.nl,
% url=https://kmitd.github.io/ilaria/,
% \address[3]{Vrije Universiteit Amsterdam, De Boelelaan 1105, 1081 HV Amsterdam, The Netherlands}

% \author[4]{Manfred Jeusfeld}[%
% orcid=0000-0002-9421-8566,
% email=Manfred.Jeusfeld@acm.org,
% url=http://conceptbase.sourceforge.net/mjf/,
% \address[4]{University of Skövde, Högskolevägen 1, 541 28 Skövde, Sweden}

\cortext[1]{Corresponding author.}
% \fntext[1]{These authors contributed equally.}

\begin{abstract}
Our study introduces an automated system leveraging large language models (LLMs) to assess the effectiveness of five key tutoring strategies: 1. \textit{giving effective praise}, 2. \textit{reacting to errors}, 3. \textit{determining what students know}, 4. \textit{helping students manage inequity}, and 5. \textit{responding to negative self-talk}. Using a public dataset from the Teacher-Student Chatroom Corpus, our system classifies each tutoring strategy as either being employed as desired or undesired. Our study utilizes GPT-3.5 with few-shot prompting to assess the use of these strategies and analyze tutoring dialogues. The results show that for the five tutoring strategies, True Negative Rates (TNR) range from 0.655 to 0.738, and Recall ranges from 0.327 to 0.432, indicating that the model is effective at excluding incorrect classifications but struggles to consistently identify the correct strategy. The strategy \textit{helping students manage inequity} showed the highest performance with a TNR of 0.738 and Recall of 0.432. The study highlights the potential of LLMs in tutoring strategy analysis and outlines directions for future improvements, including incorporating more advanced models for more nuanced feedback.

\end{abstract}

\begin{keywords}
  Qualitative Analysis \sep
  Large Language Models \sep
  Dialogue Analysis \sep
  Feedback
\end{keywords}

\maketitle

\section{Background}
\begin{figure}
  \centering
  \label{fig:system}
  \includegraphics[width=\linewidth]{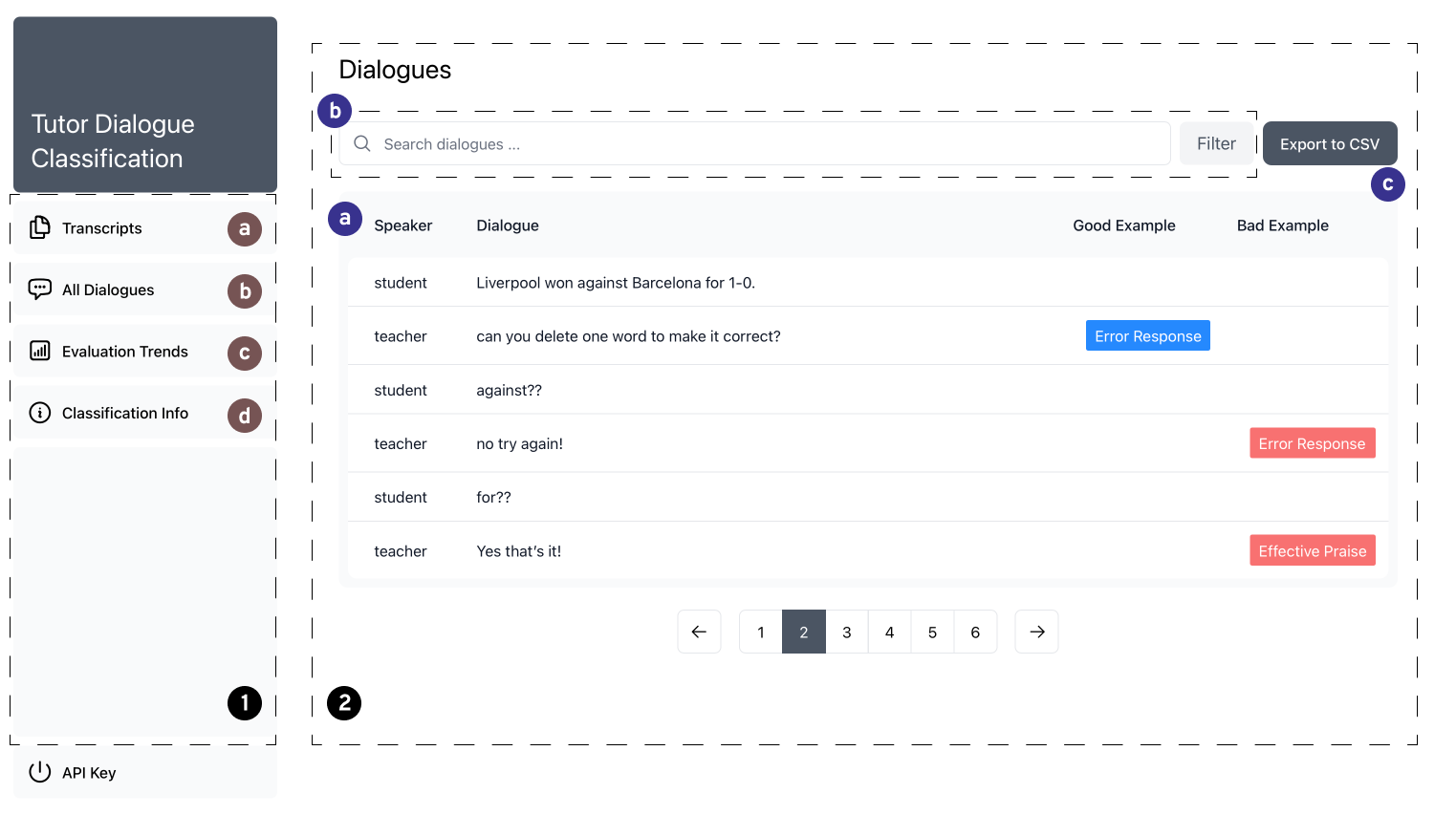}
  \caption{The Tutor Dialogue Classification system interface. On the left is (1) the navigation panel, which includes options for (a) accessing dialogue records for each transcription file, (b) viewing a comprehensive table of classification results for all available dialogue data, (c) evaluating classification patterns, and (d) explaining response categories. On the right is (2) the dialogue table, which features: (a) classification results displayed within the conversation between a tutor and a student, (b) filtering options for classification analysis, and (c) export functionalities for data analysis.}
\end{figure}

Tutoring is widely recognized as one of the most effective forms of personalized learning support \cite{NBERw27476,han2024improvingassessmenttutoringpractices}. Within tutoring sessions, strategies such as praising student effort and providing feedback play a critical role in enhancing student learning outcomes \cite{k12review,LIN2022194}. When effectively employed, these strategies can support students’ cognitive development, meet their emotional needs, and foster a positive learning environment. For example, a well-placed praise such as “\textit{You are making great progress on this problem}” (rather than generic praise like “\textit{Good job}”) can emphasize the importance of the learning process, building student resilience and motivation \cite{k12review}. Understanding how these tutoring strategies are employed during sessions is crucial, as it highlights whether they align with desired practices and are delivered in a manner that promotes student growth \cite{han2024improvingassessmenttutoringpractices}. However, the ability to automate this analysis has been constrained by the limitations of earlier natural language processing (NLP) tools, leaving room for significant improvements. Recent advancements in large language models (LLMs) offer a promising opportunity to develop automated systems for analyzing tutoring dialogues. These models (e.g., ChatGPT and Llama), with their ability to process and understand complex language patterns, provide a promising avenue for evaluating tutoring strategies in a nuanced and context-aware manner. To analyze the dialogue transcripts, our study leverages LLMs to develop an automated system (Figure \ref{fig:system}), accessible via \url{https://tutor-dialogue.vercel.app/dashboard/transcripts}.

The system is designed to detect the use of tutoring strategies and assess whether they are employed in their desired form. It allows users to upload a spreadsheet containing dialogue transcripts, with each line of dialogue and its corresponding speaker specified. As shown in Figure 1, for each strategy detected, the system determines whether it was used effectively (good) or ineffectively (bad), and this information is presented in a color-coded format for easy interpretation: blue indicates effective use (\colorbox[HTML]{A4C2F4}{good example}), while red indicates ineffective use (\colorbox[HTML]{F3CCCC}{bad example}).

\section{Method}

\subsection{Data}
Our study used the dataset provided from the Teacher-Student Chatroom Corpus \cite{caines-etal-2022-teacher}. The dataset contains one-on-one English lessons in an online chatroom. It was released in 2022, and contains a total of 262 transcriptions. Then, we hired 4 annotators to annotate a total of 9 transcriptions for usage of 5 different tutoring strategies. In our annotation scheme, we assigned the following labels to each instance: <-1> when the tutoring strategy was not applicable, <0> when the tutoring strategy was used undesirably, and <1> when the tutoring strategy was used by the tutor in a desired manner.

\subsection{Prompt Engineering}
We used few-shot chain-of-thought prompting for each of the five tutoring strategies: (1) \textit{Giving Effective Praise}, (2) \textit{Reacting to Errors}, (3) \textit{Determining what students know}, (4) \textit{Helping Students Manage Inequity}, and (5) \textit{Responding to Negative Self-Talk}. These tutoring strategies generally encourage students to persevere and increase their engagement, which are drawn from the PLUS Tutors Platform, \url{https://www.tutors.plus/en/solution/training}.

\section{Results}
Our study used the GPT-3.5 model to detect and classify tutoring strategies through few-shot prompting. Table \ref{tab:result} presents the accuracy of GPT-3.5 in identifying and classifying five tutoring strategies, measured by True Negative Rate (TNR) and Recall. GPT-3.5 achieves moderate TNR (0.655-0.738) but lower Recall (0.327-0.432). This suggests that the model performs somewhat effectively at excluding incorrect classification, but still struggles with identifying the correct one from the remaining two options. “\textit{Helping Students Manage Inequity}” strategy achieves the highest performance with TNR of 0.738 and Recall of 0.432, though overall performance remains limited.

\begin{table*}
  \caption{The performance of prompting GPT-3.5 model on identifying desired or undesired tutoring strategies}
  \label{tab:result}
  \begin{tabular}{ccl}
    \toprule
    \textbf{Tutoring Strategies} & \textbf{True Negative Rate} & \textbf{Recall} \\
    \midrule
    \textit{Giving Effective Praise} & 0.655 & 0.327\\
    \textit{Reacting to Errors} & 0.683 & 0.376\\
    \textit{Determining What Students Know} & 0.694 & 0.413\\
    \textit{Helping Students Manage Inequity} & 0.738 & 0.432\\
    \textit{Responding to Negative Self-Talk} & 0.665 & 0.331\\
  \bottomrule
\end{tabular}
\end{table*}

Further enhancements to our transcription analysis system will focus on incorporating more advanced LLMs, providing detailed statistics and feedback based on the classification results, reporting the frequency with which each strategy was used effectively or ineffectively and generating overall feedback from the model. This feedback will evaluate the tutor's effectiveness in employing each strategy and offer suggestions for improvement.

\begin{acknowledgments}
This research was funded by the Richard King Mellon Foundation (Grant \#10851) and the Learning Engineering Virtual Institute (\href{https://learning-engineering-virtual-institute.org/}{https://learning-engineering-virtual-institute.org/}). The opinions, findings, and conclusions expressed in this paper are those of the authors alone. 
\end{acknowledgments}

%% The declaration on generative AI comes in effect
%% https://ceur-ws.org/GenAI/Policy.html
\section*{Declaration on Generative AI}
  % {\em Either:}\newline
  % The author(s) have not employed any Generative AI tools.
  % \newline
  
 % \noindent{\em Or (by using the activity taxonomy in ceur-ws.org/genai-tax.html):\newline}
During the preparation of this work, the authors used GPT-4 for grammar and spelling checks. After using GPT-4, the authors reviewed and edited the content as needed and take full responsibility for the final publication.

%%
%% Define the bibliography file to be used
\bibliography{reference}

% \appendix


% The sources for the ceur-art style are available via
% \begin{itemize}
% \item \href{https://github.com/yamadharma/ceurart}{GitHub},
% % \item \href{https://www.overleaf.com/project/5e76702c4acae70001d3bc87}{Overleaf},
% \item
%   \href{https://www.overleaf.com/latex/templates/template-for-submissions-to-ceur-workshop-proceedings-ceur-ws-dot-org/pkfscdkgkhcq}{Overleaf
%     template}.
% \end{itemize}

\end{document}